\documentclass[pre,amssymb,amsmath,floats,twocolumn,superscriptaddress]{revtex4}
\usepackage{epsfig}
\usepackage{textcomp}

\begin{document}

\title{Frequency~of~occurrence~of~numbers~in~the~World~Wide~Web}

\author{Sergey N. Dorogovtsev}
\affiliation{Departamento de F{\'\i}sica da Universidade de Aveiro, 3810-193 Aveiro, Portugal}
\affiliation{A. F. Ioffe Physico-Technical Institute, 194021
  St. Petersburg, Russia}

\author{Jos\'e Fernando F. Mendes}
\affiliation{Departamento de F{\'\i}sica da Universidade de Aveiro, 3810-193 Aveiro, Portugal}

\author{Jo\~ao Gama Oliveira\footnote{To whom correspondence 
should be addressed. E-mail: joao.gama@nd.edu}}
\affiliation{Departamento de F{\'\i}sica da Universidade de Aveiro, 3810-193 Aveiro, Portugal}
\affiliation{Department of Physics and Center for Complex Network Research, University of Notre Dame, IN 46556, USA}

\date{}

\begin{abstract}

The distribution of numbers in human
documents is determined by a variety of diverse natural and human
factors, whose relative significance can be evaluated by studying
the numbers' frequency of occurrence. Although it has been studied
since the 1880's,  
this subject remains
poorly understood. Here, we obtain the detailed statistics of
numbers in the World Wide Web, finding that their distribution is a
heavy-tailed dependence which splits in a set of power-law ones. 
In
particular, we find that the frequency of numbers associated to
western calendar years shows an uneven behavior: 2004 represents a
`singular critical' point, appearing with a strikingly high
frequency; as we move away from it, the decreasing frequency allows
us to compare the amounts of existing information on the past and on
the future. Moreover, while powers of ten occur extremely often,
allowing us to obtain statistics up to the huge $10^{127}$,
`non-round' numbers occur in a much more limited range, the
variations of their frequencies being dramatically different from
standard statistical fluctuations. 
These findings provide a view of
the array of numbers used by humans as a highly non-equilibrium and
inhomogeneous system, and shed a new light on an issue that, once
fully investigated, could lead to a better understanding of many
sociological and psychological phenomena.

\end{abstract}


\maketitle

\newpage

Already in the early 1880's, Newcomb \cite{n881} noticed a specific
uneven distribution of the first digits of numbers, which is now
known as Benford's law \cite{b38}. The observed form of this
distribution indicates the wide, skewed shape of the frequency of
occurrence of numbers in nature 
[3--5] ---for
illustration, and to clarify the question, note that in these first two
sentences the numerals 1, 2, 3, 5 and 1880 all occur twice. Benford's
law is directly derived by assuming that a number occurs with a
frequency inversely proportional to it, meaning that the frequencies
of numbers in the intervals $(1,10)$, $(10,100)$, $(100,1000)$, etc.
are equal. Yet, this assumption lacks convincing quantitative support
and understanding, in part due to scanty data available. In our days,
this problem can be tackled by resorting (with the help of search
engines) to the enormous database constituted by the World Wide Web.

One should note that the profoundly wide form of the distribution of
numbers in human documents is determined by two sets of factors. The
first includes general natural reasons of which the most important
is the multi-scale organization of our World. The second are `human
factors' including the current technological level of the society,
the structure of languages, adopted numeral and calendar systems,
history, cultural traditions and religions, human psychology, and
many others. By analyzing the occurrence frequency of numbers we can
estimate the relative significance and role of these factors.


\begin{table}[b]
\caption{
\label{t1}
{\bf Typical numbers with high frequencies of occurrence}
}
\begin{ruledtabular}
\begin{tabular}{cl}
Example & Description\\
\hline
1000 &  powers of 10
\\
2460, 2465 & `round' numbers: multiples of 10 and 5
\\
666,\footnotemark[1] 131313 & numbers easy to remember or symmetric
\\
$512 = 2^9$ & powers of 2
\\
666,\footnotemark[1] 777 & numbers with strong associations
\\
78701 & popular zip codes
\\
866, 877 & toll free telephone numbers
\\
1812 & important historical dates
\\
747, 8086 & serial numbers of popular products
\\
314159 & beginning parts of mathematical constants
\\
\end{tabular}
\end{ruledtabular}
\footnotetext[1]{A number may occur simultaneously in several lines of the table.}
\end{table}


The frequency of occurrence of numbers in the World Wide Web pages
(or, in other words, WWW documents) necessarily reflects the
distribution of numbers in all human documents, allowing us to
effectively study their statistics by using search engines, which
usually supply the approximate number of pages containing the Arabic
numeral that we are looking for. In this respect, the WWW
provides us with huge statistics. Yet, the frequencies of occurrence
of distinct kinds of numbers are very different \cite{lwf02}: for
example, one can see that 777 and 1000 occur much more frequently
than their neighbors (Table \ref{t1}). Here we report on the
markedly distinct statistics of different types of natural numbers
(or, rather, positive integers) in the WWW documents, collected
through the currently most popular search engine \cite{g}. We
consider separately (i) powers of 10 and (ii) non-round integers,
and find that in both of these cases, the number $N(n)$ of pages
containing an integer $n$ decays as a power law, $N(n) \sim
n^{-\beta}$, over many orders of magnitude. The observed values of
the $\beta$ exponent strongly differ for the different types of
numbers, (i) and (ii), and also differ from 1, thus contradicting
the above mentioned assumption of inverse proportionality for their
frequency of occurrence.

Note that previously scale-free (i.e. power-law) distributions
were observed for processes in the WWW \cite{hppl98,ha99} and its
structural characteristics \cite{ajb99,ba99}. However, and in
contrast to these studies, we use the WWW as a database for
measuring one of the basic distributions in nature. In order to
explain the observed distributions, we treat the global array of
numbers as a non-equilibrium, evolving system with a specific influx
of numbers, and, as a reflection of this non-equilibrium nature, we
find a `critical behavior' of $N(n)$ in the neighborhood of $n=2004$
(the current year at the time the measurements were made): near this
point, the frequency of WWW documents follows a power law, $N(n)
\sim (2005-n)^{-\alpha}$.


\begin{figure}[t]
\begin{center}
\scalebox{0.32}{\includegraphics{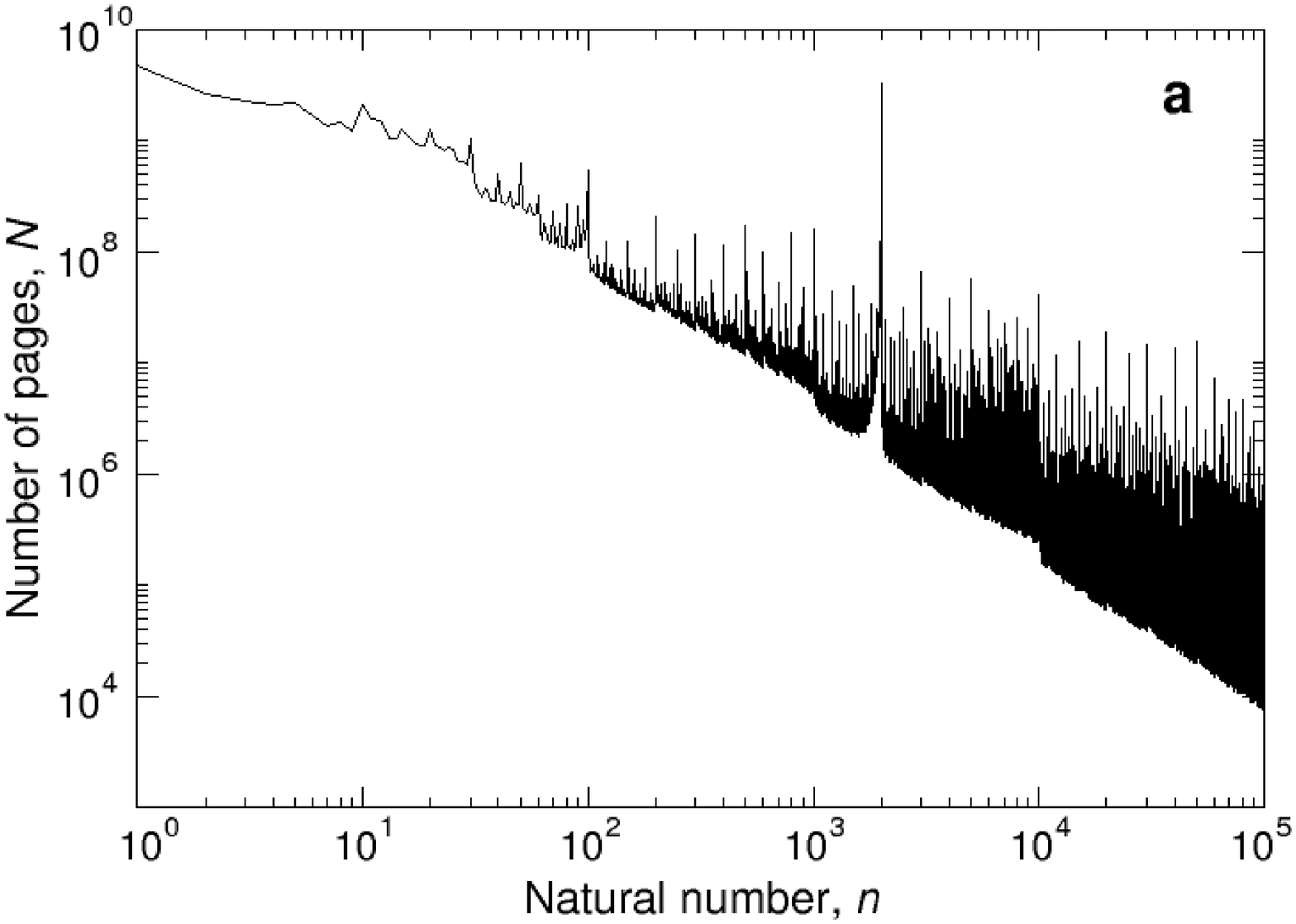}}
\scalebox{0.1908}{\includegraphics{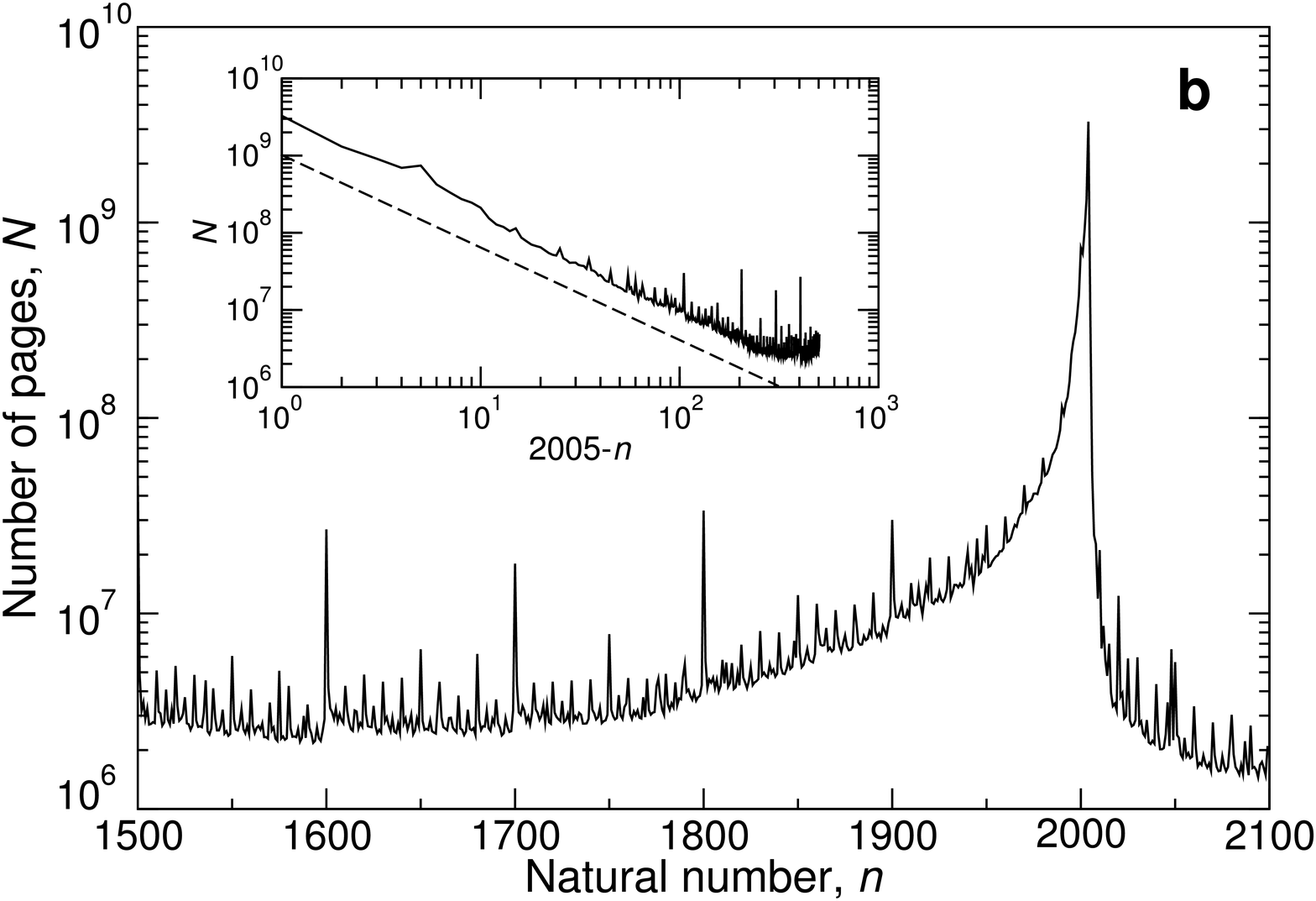}}
\end{center}
\caption{ {\bf a,} The frequencies $N(n)$ of WWW pages containing
numbers $n$ up to $100,\!000$ on a log-log plot. Note the peak at
$n=2004$. {\bf b,} The part of the distribution around $n=2004$
shown in more detail on a log-linear plot. The asymmetric form of
the peak gives an idea about the relation between the stored volumes
of information on the past and on the future: the former is much
more referred to than the latter. In the inset, the low-$n$ part of
this peak is plotted versus the difference $2005-n$ on a log-log
plot ($1500 < n < 2005$). A power-law behavior is observed
practically in the entire range where the contribution of numbers
associated to years is main. The slope of the dashed line is $-1.2$.
It was not possible to find a reliable fit to the dependence for $n
\geq 2005$. These plots also demonstrate a hierarchy of peaks for
documents holding numbers of different kinds. } \label{f1}
\end{figure}


Finally, we show that the statistics of variations of the
frequencies of WWW pages which contain close numbers of the same
kind, dramatically disagrees with the standard distribution of
statistical fluctuations. We observe, namely, that the amplitude of
these variations, $\delta N(n)$, is much greater than what would be
expected for standard statistical fluctuations. Consequently, the
frequencies of pages containing different numbers fluctuate not
independently, these fluctuations being a reflection of those of the
influx of numbers.

\section*{Current-year Singularity}

In the second week of December 2004, we obtained the frequency of
WWW documents corresponding to positive integers $n$ in the range
between 1 and $100,\!000$ (Fig.~\ref{f1}a). This plot contains a set
of regularly distributed peaks, which indicate that different types
of numbers occur with very unlike frequencies. For example, the
number of documents containing round (ending with 0) numbers is much
higher than that for non-round numbers. Furthermore, the special
number 2004 occurs with a remarkably high frequency:
$3,\!030,\!000,\!000$ pages. For comparison, among
$8,\!058,\!044,\!651$ WWW pages covered by the used search engine, a
single character string $a$ occurs in about $8,\!000,\!000,\!000$
pages, while the numbers 0, 1 and 1000 occur in
$2,\!180,\!000,\!000$, $4,\!710,\!000,\!000$ and $154,\!000,\!000$
pages, respectively. The high, asymmetric peak of $N(n)$ around
$n=2004$ (Fig.~\ref{f1}b) is naturally identified as the
contribution of documents containing numbers associated to years;
below $n=2005$, this peak can be fitted by a power law, following
$N(n) \sim (2005-n)^{-\alpha}$, where $\alpha = 1.2 \pm 0.1$ (inset
of Fig.~\ref{f1}b). Therefore, in the vicinity of 2004, $N(n)$
increases with $n$ much faster than the total number of pages in the
WWW grows with time, which indicates that there are many pages with
numbers associated to years that disappear from the WWW (or at
least, are updated) after a while. Indeed, our observations prove
that the amount of pages holding a number $n<t$ (where $t$ is time
measured in years) in the region of the `critical singularity'
decreases with $t$ approximately following $N(n,t) \sim
(t-n)^{-\alpha}$. 


\begin{figure}
\begin{center}
\scalebox{0.178}{\includegraphics{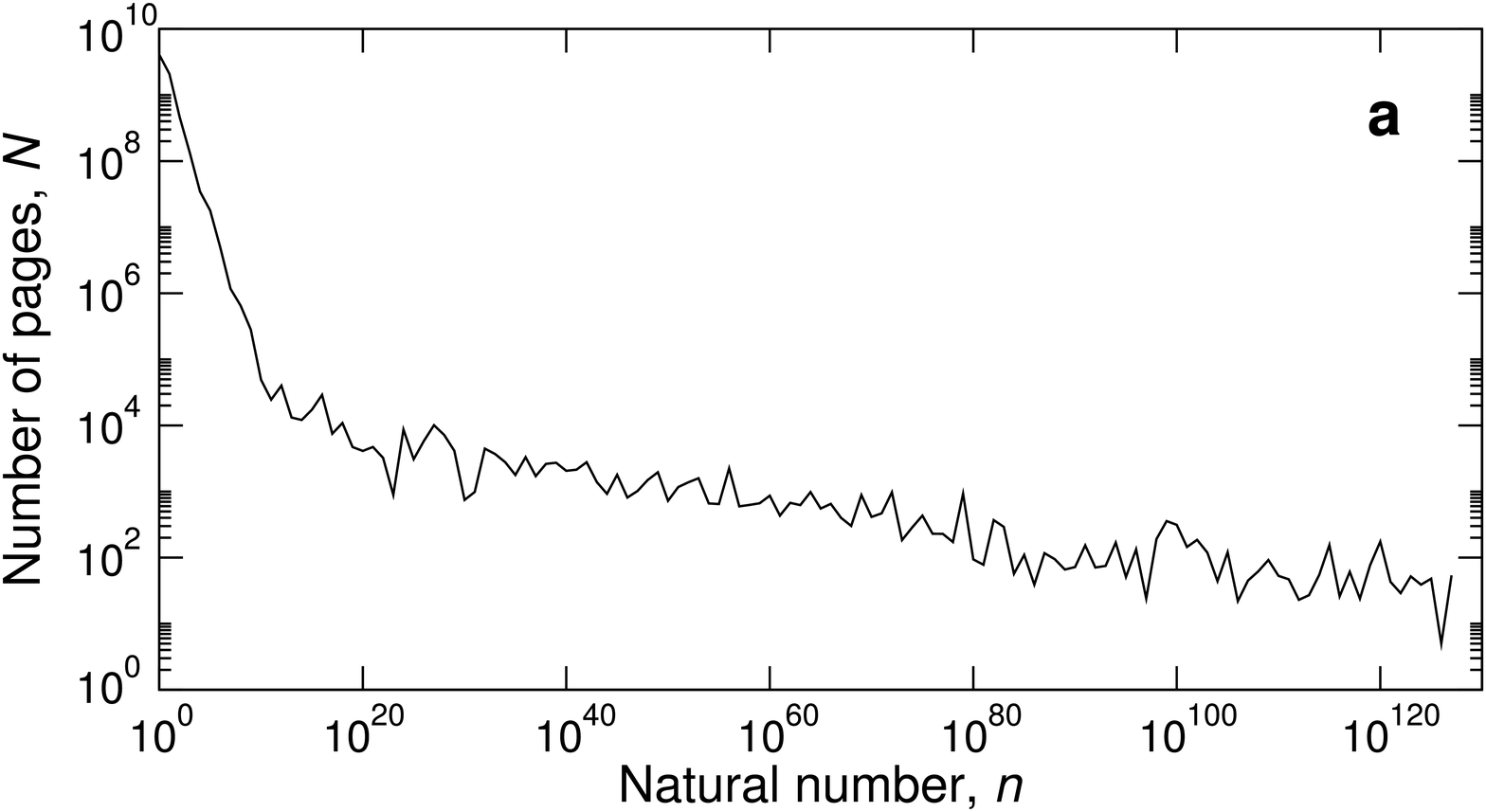}}
\scalebox{0.178}{\includegraphics{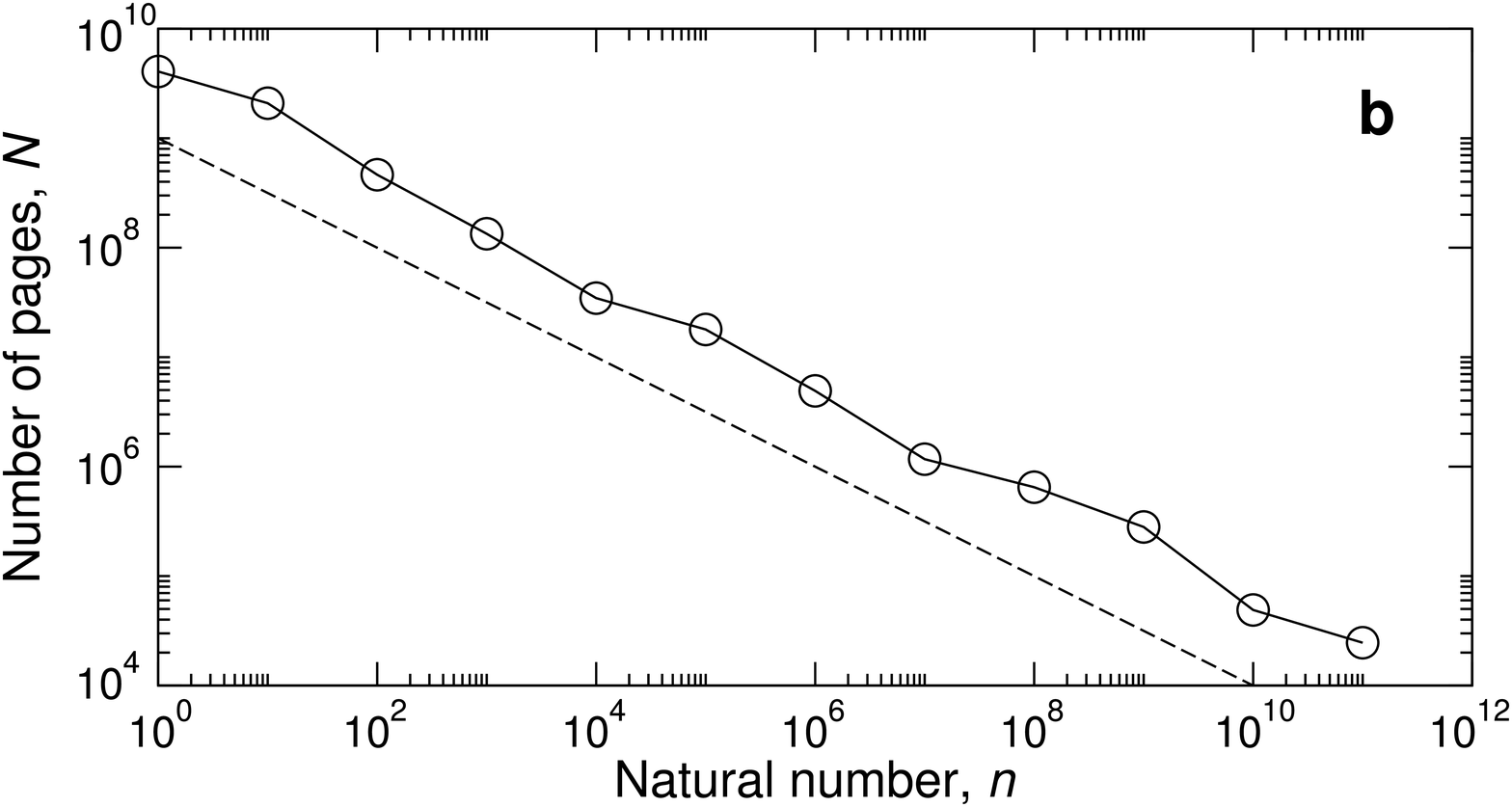}}
\scalebox{0.178}{\includegraphics{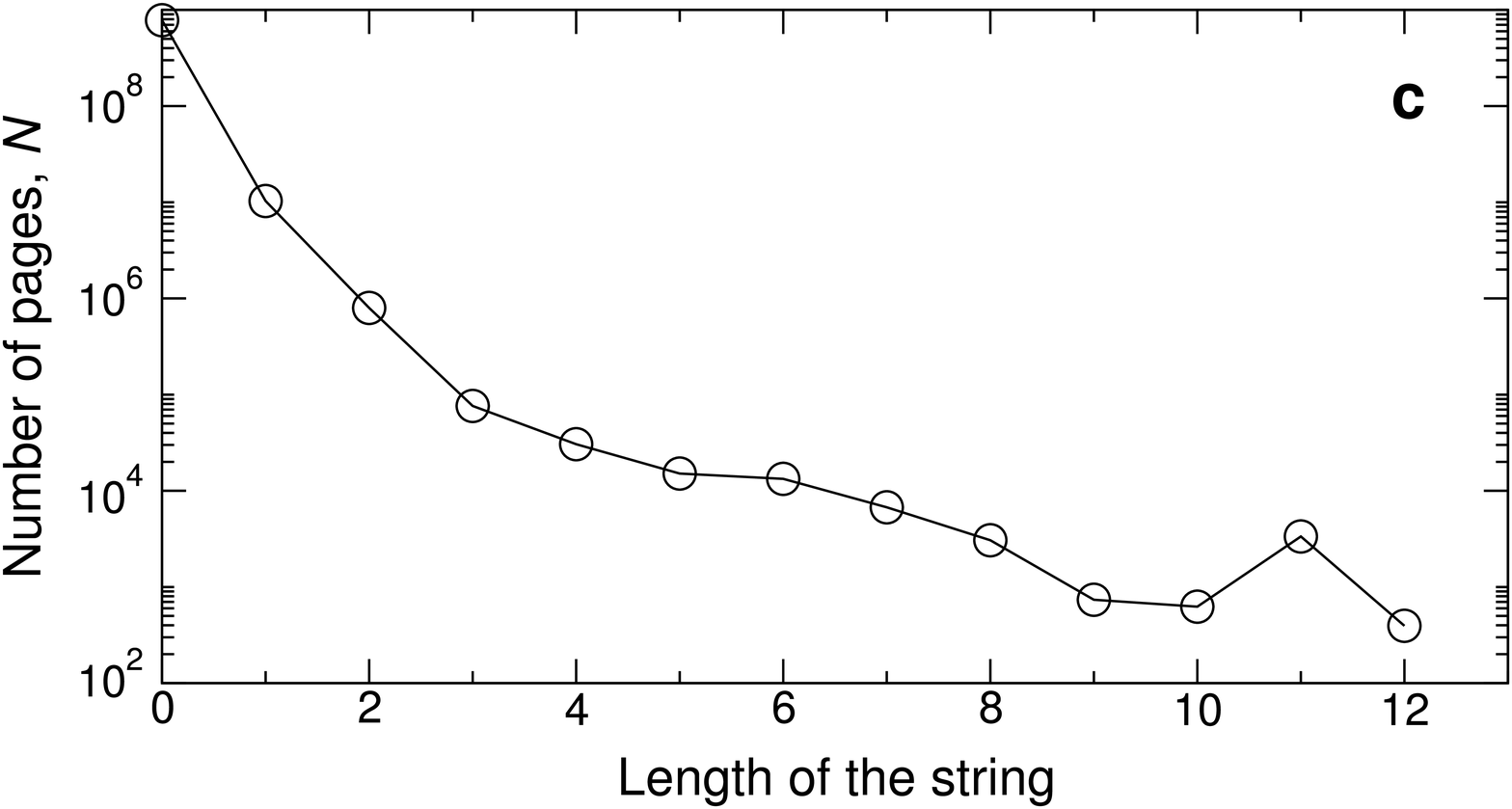}}
\end{center}
\caption{ The frequencies of WWW pages containing powers of $10$.
{\bf a,} The full log-log plot up to the maximal searchable
$10^{127}$. {\bf b,} The power-law-like part of the distribution.
The slope of the dashed line is $-0.5$. We emphasize that the
power-law dependence is observed over $11$ orders of magnitude,
which is a uniquely wide range. {\bf c,} For comparison, the number
of WWW documents containing a character string {\em baaa\!
\ldots\!a} of varying length on a log-linear plot (the length of the
string is the equivalent to the exponent in the power of $10$). Note
the difference from {\bf b}. } 
\label{f2}
\end{figure}


\section*{Power-law Distributions}

We find that the frequency of occurrence of natural numbers,
considered without separating them into distinct classes
(Fig.~\ref{f1}a), is a slowly decreasing dependence. Nevertheless,
it can hardly be fitted by any power law because it is, in fact, the
result of the superposition of distributions of distinct kinds of
numbers, which, in turn, are power laws having different exponents.
In order to proceed, we then compare the statistics of the WWW
documents which hold two `extreme' types of numbers: (i) powers of
10, which should occur with the highest frequencies due to the common
decimal numeral system, and, contrastingly, (ii) non-round numbers
(i.e. those with a non-zero digit in the end) which are, on average,
the most indistinctive ones, therefore occurring with the lowest
frequencies. It is worth remarking that, even though the non-round
include many peculiar numbers, such as 777 for example, we find
that their contribution does not change the statistics noticeably.


\begin{figure}
\begin{center}
\scalebox{0.195}{\includegraphics{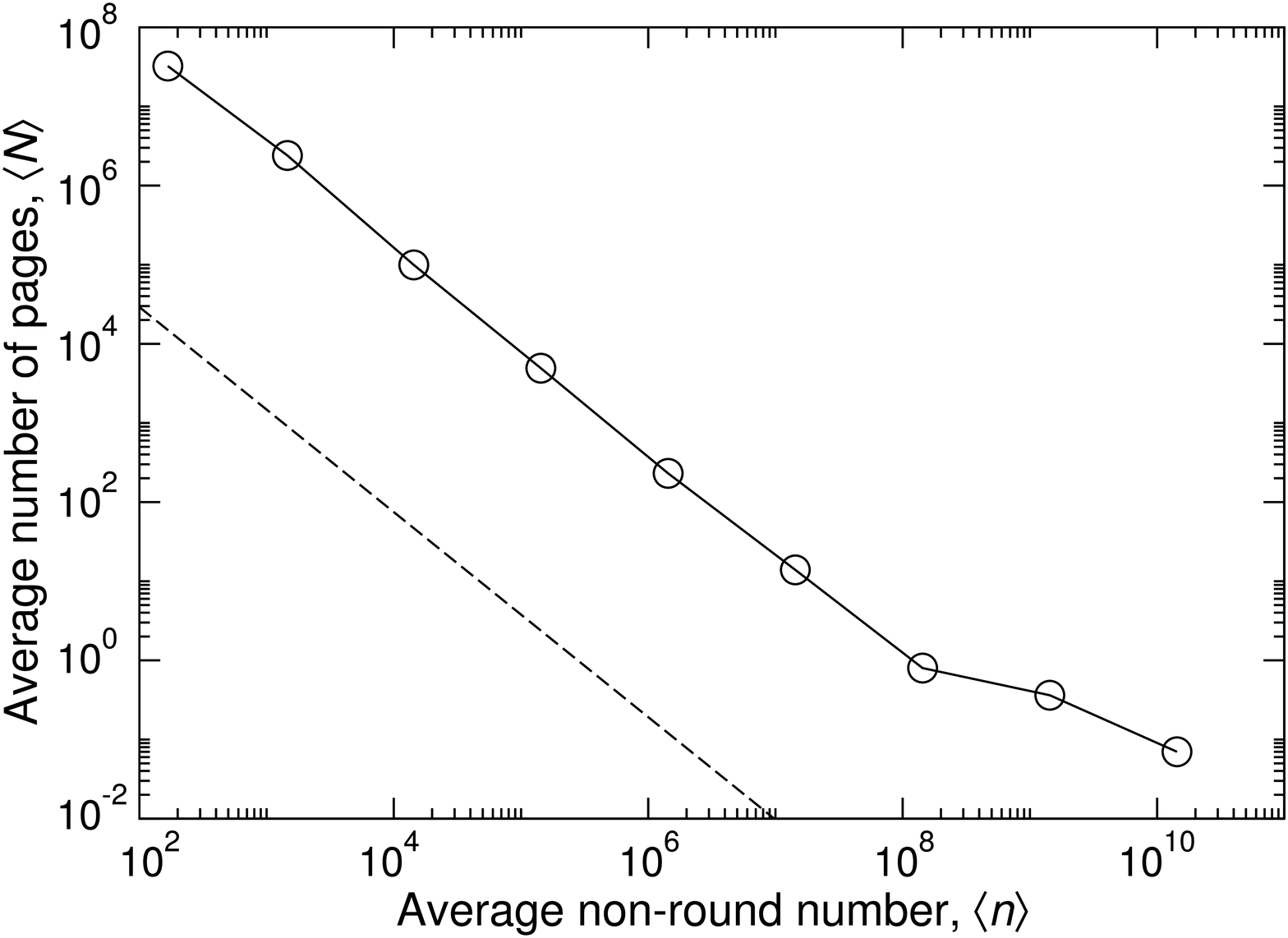}}
\end{center}
\caption{ Log-log plot of the frequencies of WWW pages holding
non-round numbers. The circles show the average amounts of pages
with non-round numbers taken from relatively narrow intervals ($50$
numbers). Each interval is centered at the $\langle n \rangle$
coordinate of a circle. The dashed line has slope $-1.3$. Note that
the power-law behavior is observed over $6$ orders of magnitude.
Non-round numbers occur much less frequently than powers of 10,
which explains the essentially narrower range of numbers in this
plot than in Fig. \protect\ref{f2}a. For instance, presently, and as
far as search engines report, there are no WWW documents with the
number 12345789013.
}
\label{f3}
\end{figure}


The strikingly high frequency of occurrence of powers of 10 in the
WWW allows us to obtain the statistics for numbers up to $10^{127}$
(Fig.~\ref{f2}a), a range that is restricted by the limited size of
strings being accepted by the used search engine (128 characters).
Two distinct regions are seen in the distribution. The
region of relatively `small' numbers, up to $10^{11}$ (Fig.~\ref{f2}b),
is of a power-law form, $N(n) \sim n^{-\beta}$, where
$\beta=0.50\pm0.02$, hence close to the law $N(n) \sim 1/\sqrt{n}\,$;
note that this exponent is much smaller than 1 and far smaller than
the values of the exponents of typical Zipf's law distributions
\cite{ba99,z49}, these being mostly in the range between 2 and 3.
For comparison, the occurrence frequencies of a character string
{\em baaa\! \ldots\!a} of varying length were also measured, a quite
different, far from straight line, dependence having been observed
(Fig.~\ref{f2}c). For $n$ larger than $10^{11}$, we observe an
extremely slow decrease of the frequency of occurrence of pages
containing powers of $10$ (Fig. 2a). It is worth noting that the
crossover between these two regimes turns out to be rather close to
the maximum 32 digit binary number, which is about $0.4 \times
10^{10}$. 

For properly measuring the occurrence frequency of
non-round numbers, we use a set of intervals selected in their wide
range, each of which having a width of 50 numbers, so that the
relative variation of the frequency of WWW pages inside a specific
interval is sufficiently small. In addition, these intervals are
chosen far from the powers of 10, whose close neighborhood includes
numbers, such as, for instance, 1009, that occur more often and
whose distribution does not follow a clear power law. Within each of
these intervals, we take the average values of $n$ and $N(n)$, and
denote them by $\langle n \rangle$ and $\langle N \rangle$,
respectively; the resulting dependence (Fig.~\ref{f3}) has a
prominent power-law region with exponent $\beta=1.3\pm0.05$, which
strongly differs from that ascertained for powers of 10. As numbers
grow, the ratio of the amount of WWW documents with powers of 10 to
that with non-round numbers increases, following the $n^{0.8}$
dependence.


\begin{figure}[t]
\begin{center}
\scalebox{0.195}{\includegraphics{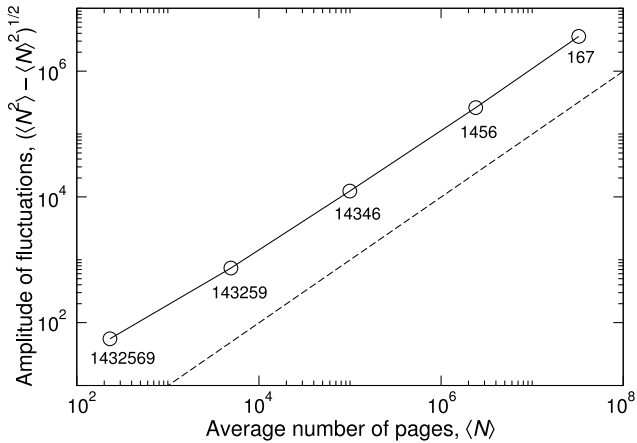}}
\end{center}
\caption{ Amplitude of the fluctuations, $\sqrt{\langle N^2 \rangle
- \langle N \rangle^2}$, of the frequencies of pages containing
non-round numbers versus their mean values, $\langle N \rangle$, on
a log-log plot. The data (circles) were obtained resorting to the
same intervals as in Fig.~\protect\ref{f3}. Next to each circle the
average (non-round) number, $\langle n \rangle$, for the
corresponding interval is indicated. The dashed line has slope $1$.
One can see that $\sqrt{\langle N^2 \rangle - \langle N \rangle^2}
\approx 0.1\langle N \rangle$ for $\langle N \rangle>10^3$. }
\label{f4}
\end{figure}


A few mechanisms generating power-law distributions \cite{z49} are
known \cite{s55,y25,w22,m77,b96}. Most of these mechanisms explain
power laws as a result of a specific self-organization of a
non-equilibrium system, and we treat our observations in the spirit
of these approaches. Evidently, the array of numbers in human
documents is an evolving system, and the stochastic growth of this
array is due to a permanent influx of numbers, added with new
documents. The added numbers (among which may also occur new
distinct ones, that were not employed previously) are chosen from a
distribution which is determined by the one for the existing
numbers. Here we do not discuss a specific model exploiting this
mechanism and generating the observed complex distributions,
but, instead, we explain the reason for the unusual
small values of exponents which we observed --- $\beta=0.5$ and
$1.3$ (Figs.~\ref{f2}b and \ref{f3}), while typical Zipf's law
exponents are $2$ and greater. At least, Zipf's law exponents must
take values greater than $1$. At first sight, this difference seems
surprising, since the mechanisms of the power laws are quite
similar. But, importantly, these two sets of
exponents are defined for different distributions. In our
non-traditional case, the observed power law describes the behavior
of the frequency of WWW pages with a given natural number $n$,
namely $N(n) \sim n^{-\beta}$. In contrast, typical Zipf's law
exponent $\gamma$ occurs in a power law for a quite different
quantity: in our terms, this quantity is the amount, $m(N)$, of
distinct numbers, where each of them occurs in every of $N$
Web pages. So, we have the relation $m(N) \sim N^{-\gamma}$. One can
show that the exponents $\beta$ and $\gamma$ satisfy a simple
relation, $\beta=1/(\gamma-1)$ \cite{dmbook03}. As a result, if the
$\gamma$ exponent is greater than $2$, which is typical for simple
linear growth processes, the $\beta$ exponent is smaller than $1$,
as in Fig.~\ref{f2}b. On the other hand, nonlinear growth may produce
exponents $\gamma$ below $2$, which gives $\beta$ greater than $1$,
as in Fig.~\ref{f3}.

\section*{Fluctuations of the Number of WWW Pages}

The distributions reported here demonstrate that the frequencies of
WWW pages holding numbers even of the same kind (for example,
non-round numbers) strongly fluctuate from number to number. For
documents containing non-round integers, we obtain the dependence of
the fluctuations' amplitude (i.e. dispersion), $\sqrt{\langle (N -
\langle N \rangle)^2 \rangle} = \sqrt{\langle N^2 \rangle - \langle
N \rangle^2}$, on the average frequency, $\langle N \rangle$, of
these documents (Fig.~\ref{f4}). For calculating these dispersions
and mean values, we used the same intervals as in Fig.~\ref{f3}. The
resulting dependence turns out to be proportional, $\sqrt{\langle
N^2 \rangle - \langle N \rangle^2} \approx 0.1 \langle N \rangle$,
over a broad region of values $\langle N \rangle$, which crucially
differs from the square root behavior of standard statistical
fluctuations \cite{ll93}. The usual reason for such a strong
difference is that the fluctuations of the quantities under study
are not statistically independent \cite{mb04,mb04(2)}. In this respect,
there is only one factor in the evolution of the array of numbers
which can break the statistical independence of fluctuations,
namely, the variation of the influx of numbers. So, the observed
proportional law proves that the variations of the occurrence
frequencies of numbers are an outcome of the fluctuations of their
global influx in the WWW.

\section*{Discussion and Conclusions}

These observations suggest a new view of the array of integers in the
WWW (and in nature) as a complex, evolving, inhomogeneous system.
The statistics of numbers turns out to be far more rich and complex
than one might expect based on classical Benford's law. Moreover,
our findings provide a tool for extracting meaningful information
from statistical data on the frequency of occurrence of numbers. As
an illustration, consider the two integers, 666 and 777, with clear
associations. We find that these numbers occur in the WWW with
frequencies of $11,\!800,\!000$ and $13,\!600,\!000$ pages,
respectively, which are $1.25$ and $1.65$ times higher than, on
average, the occurrence frequencies of their non-round neighbors.
These deviations are to a great extent higher than what one would
anticipate from the relative amplitude of fluctuations, 0.1. Therefore,
we can reasonably compare the amounts of pages containing 666 and 777
obtained after subtracting the numbers of pages holding the
neighbors of these two integers. These subtractions give
$2,\!400,\!000$ and $5,\!400,\!000$ pages for $666$ and $777$,
respectively. It is the difference (or, rather, the relative
difference) between the two last amounts that should be used as a
starting point for a subsequent comparative analysis. The proposed
approach is very suggestive. Indeed, by analyzing the frequencies of
occurrence of specific `popular' numbers with clear interpretations
one could evaluate the relative significance of the corresponding
underlining factors of this popularity.

Many more questions lie ahead: How do the occurrence frequencies of
specific numbers vary in time? How do different numbers correlate
and co-occur in WWW documents? It is well known that humans can
easily memorize only up to rather limited sequences of digits
\cite{m56,c01}, which are, therefore, many times replaced by words
(like, for instance, the IP addresses of computers). Then, how does
the statistics of numbers relate to the organization of human memory
and to semantics? Our findings quantitatively show the key role of
the common decimal numeral system --- a direct consequence of the
number of fingers. How do other numeral systems (the binary system,
for example) influence the general statistics of numbers?

The global array of numbers is surmised to be a ``numeric snapshot of
the collective consciousness'' \cite{lwf02}. So, the study of their
statistics could lead to a better understanding of a wide
circle of sociological and psychological phenomena. The distribution
of numbers in human documents contains a wealth of diverse
information in an integrated form. The detailed analysis of the
general statistics of numbers in the WWW could allow the effective
extraction and evaluation of this hidden information. 

We wish to thank Albert-L\'aszl\'o
Barab\'asi for fruitful discussions, suggestions and comments on the
manuscript. We also thank Eivind Almaas and Alexei V\'azquez for
comments on the manuscript, and G\'abor Szab\'o for a useful
discussion. This work was partially supported by projects
POCTI/FAT/46241/2002 and POCTI/MAT/46176/2002. S.N.D. and J.F.F.M.
acknowledge the NATO program OUTREACH for support. J.G.O.
acknowledges financial support of FCT, grant No. SFRH/BD/14168/2003.

\end{document}